\documentclass[prb,aps,twocolumn,amsmath,amssymb,citeautoscript,superscriptaddress]{revtex4-1}

\usepackage{graphicx}% Include figure files
\usepackage{dcolumn}% Align table columns on decimal point
\usepackage{bm}% bold math
\usepackage{color}

\begin{document}

\title{Green thermoelectrics: Observation and analysis of plant thermoelectric response}

\author{Christophe Goupil}\email{christophe.goupil@univ-paris-diderot.fr}
\affiliation{Laboratoire Interdisciplinaire des Energies de Demain (LIED) UMR 8236 Universit\'e Paris Diderot CNRS 4 Rue Elsa Morante 75013 Paris France}
\author{Henni Ouerdane}
\affiliation{Laboratoire Interdisciplinaire des Energies de Demain (LIED) UMR 8236 Universit\'e Paris Diderot CNRS  4 Rue Elsa Morante 75013 Paris France}
\affiliation{Russian Quantum Center, 100 Novaya Street, Skolkovo, Moscow region 143025, Russia}
\author{Arnold Khamsing}
\affiliation{Laboratoire Interdisciplinaire des Energies de Demain (LIED) UMR 8236 Universit\'e Paris Diderot CNRS 4 Rue Elsa Morante 75013 Paris France}
\author{Yann Apertet}
\affiliation{Lyc\'ee Jacques Pr\'evert, 30 Route de Saint Paul, 27500 Pont-Audemer, France}
\author{Fran{\c c}ois Bouteau}
\affiliation{Laboratoire Interdisciplinaire des Energies de Demain (LIED) UMR 8236 Universit\'e Paris Diderot CNRS  4 Rue Elsa Morante 75013 Paris France}
\affiliation{Laboratorio Internazionale di Neurobiologia Vegetale - Department of Plant Soil \& Environmental Science, University of Florence, Florence, Italy}
\author{Stefano Mancuso}
\affiliation{Laboratorio Internazionale di Neurobiologia Vegetale - Department of Plant Soil \& Environmental Science, University of Florence, Florence, Italy}
\affiliation{Universit\'e Paris Diderot, Sorbonne Paris Cit\'e, Paris Interdisciplinary Energy Research Institute (PIERI), Paris, France}
\author{Rodrigo Pati{\~n}o}
\affiliation{Laboratoire Interdisciplinaire des Energies de Demain (LIED) UMR 8236 Universit\'e Paris Diderot CNRS  4 Rue Elsa Morante 75013 Paris France}
\affiliation{Departamento de F\'{\i}sica Aplicada, Cinvestav-Unidad M\'erida, AP 73 Cordemex, 97310 M\'erida, Yucatan, Mexico}
\author{Philippe Lecoeur}
\affiliation{Institut d'Electronique Fondamentale, Universit\'e Paris-Sud, CNRS, UMR 8622, F-91405 Orsay, France}

\date{\today}

\begin{abstract}
Plants are sensitive to thermal and electrical effects; yet the coupling 
of both, known as thermoelectricity, and its quantitative measurement in vegetal
systems never were reported. We recorded the thermoelectric response of bean sprouts 
under various thermal conditions and stress. The obtained experimental data unambiguously 
demonstrate that a temperature difference between the roots and the leaves of a bean sprout 
induces a thermoelectric voltage between these two points. Basing our analysis of the data 
on the force-flux formalism of linear response theory, we found that the strength of the 
vegetal equivalent to the thermoelectric coupling is one order of magnitude larger than 
that in the best thermoelectric materials. Experimental data also show the importance of 
the thermal stress variation rate in the plant's electrophysiological response. Therefore, 
thermoelectric effects are sufficiently important to partake in the complex and intertwined 
processes of energy and matter transport within plants.
\end{abstract}
\maketitle

\section{Introduction}
Plants are complex biological systems, which feed, develop and function thanks to a variety 
of elementary and cooperative processes that ensure transport of energy and
matter in the form of mineral elements, carbohydrates, and hormones in watery
solutions called sap. Two categories of sap are transported in
conductive tissues: xylem sap transported in xylem, a continuum of dead cells forming 
a pipe that opposes little resistance to the transport of water and mineral elements  
from roots to leaves; and phloem sap, containing water and sugar, formed in
the leaves by photosynthesis, and transported in phloem (sieve tube elements)
from the leaves to the roots. Explanations for xylem sap transport includes the
cohesion-tension theory \cite{DixonJoly,Askenasy} and multiforce theories
\cite{Tyree,Zimmerman,Wang}; and the pressure flow
hypothesis proposes a mechanism for phloem sap transport \cite{DeSchepper2013}. 
Notwithstanding their relative merits, it is of interest to note that as
saps contain ions, a net voltage may be recorded within plants. Furthermore,
as a living plant is naturally submitted to a temperature gradient between its
roots and its leaves, heat may also flow from its hotter to its colder
regions. These two facts lead to the questions of the existence and strength
of a coupling between electrical current and heat flux in plants, and whether
this coupling plays a significant role in the plants' lives. Answers to these
questions are of importance to gain further insight into plant homeostasis,
closely related to the Le Chatelier-Braun principle in chemical
thermodynamics, and to develop further the field of plant electrophysiology by
analysing the impact of thermoelectric coupling on photosynthesis, plant
respiration, and any other processes involving bioelectrochemical phenomena
\cite{PlantEP}.

Biophysical processes, which include energy and matter transport in dedicated
structures, are essentially nonequilibrium in nature. Their description
involves the notions of thermodynamic forces and their conjugate fluxes which,
in a linear response description, are proportional to each other. In a general
manner, transport phenomena are nonequilibrium irreversible
processes\cite{deGroot,Pottier2007}: fluxes within a system result from
applied external constraints, and they are accompanied by energy dissipation
and entropy production. As the forces applied to a system derive from 
potentials, the description of the system's nonequilibrium properties relies
on the definition of these potentials and their degree of coupling
\cite{KedemCaplan}. Thermoelectricity is sometimes mistakenly viewed as a 
phenomenon pertaining to solid-state systems\cite{bookTE1,bookTE2}, but this effect has 
also been reported in liquids\cite{LiquidThermoelec} and gels\cite{GelThermoelec}.
Now, considering more specifically electrical phenomena in plants, one should also 
account for their dependence on temperature and the local variations of this latter. 
As a matter of fact, one learns from the equilibrium thermodynamics of charged fluids 
that heat and electricity are fundamentally related through the Gibbs-Duhem relation; 
so, since plants, as any other system, are subjected to the laws of thermodynamics, 
one may anticipate that a proper understanding of the coupled transport of heat and 
electricity in plants will pave the way towards the potentially rich new field of 
thermoelectricity-based plant energy conversion, and harvesting metabolic energy 
from plants \cite{voltreepower}.

Thermoelectricity as part of non-isothermal processes in living organisms, has already 
been mentioned; but except for the case of insects like hornets and bees \cite{Ishay,Galushko,Volynchik}, 
for which genuine thermoeletric effects have been analysed, the other studies simply reported the 
record of the temperature rise of Colocasia odora leaves using a thermocouple \cite{VanBeek}, or analysed  
pyroelectricity, mistakenly taken as a thermoelectricity, amongst the nonequilibrium processes 
at the origin of life \cite{Muller}. We propose here to consider bean sprouts as an illustrative case of a 
living plant system whose physiological properties also entail thermoelectric effects, with a particular focus 
on the thermovoltage response of the plant. While a voltage
difference along the stem of a plant has already been measured and analysed \cite{Love},
its interpretation does not account for a likely the thermoelectric origin. Nevertheless, 
it is clear that the source of any voltage difference 
is due to concentration gradients of charges inside the plant, regardless of the cause 
of the gradients. The underlying processes may thus originate in various electrochemical 
processes, including acid-base and redox effects\cite{Love}. In the present work we
focus on the voltage response due to a temperature difference. The system is
presented in Fig. 1. The stem and leaves are kept at constant temperature while the
temperature of the roots can be modified; the thermoelectric response is
recorded on both the plant (a) and a control wire (b).

\begin{figure}[th]
\includegraphics*[width=0.5\textwidth]{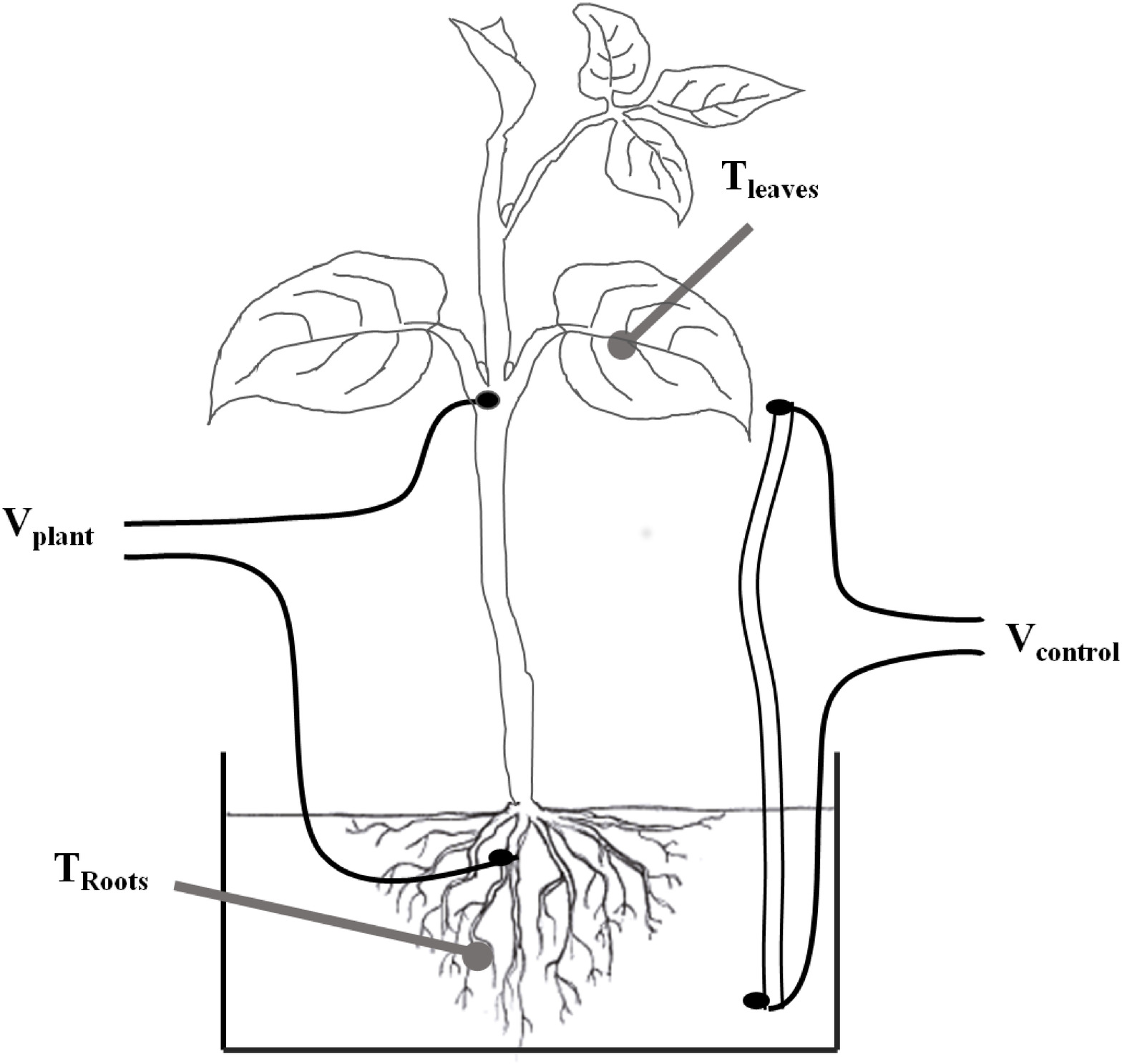} \caption{General view of
the experimental setup. The voltage records are respectively on the plant (a),
and the control wire (b).}
\label{fig:figure1}
\end{figure}

\section{Theoretical considerations}

Thermoelectricity differs from electrochemistry as the latter considers
isothermal systems only. Thermoelectric effects, on the contrary, manifest
themselves in non-isothermal conditions and their thermodynamic description is
best done with the coupled variables $(\mu,T)$, where $\mu$ is the
electrochemical potential and $T$ the temperature of the considered system. 
In a force-flux approach, as a temperature bias is
applied to a conductor, both heat and charges are transported in coupled flows
since each electron carries an electric charge, $e$, and energy; the thermoelectric 
coupling parameter or Seebeck coefficient $\alpha$ is thus defined as: 

\begin{equation}
\label{Seebeck}\alpha\overrightarrow{\nabla} T= - \frac{1}{e}%
\overrightarrow{\nabla} \mu
\end{equation}

\noindent where $\overrightarrow{\nabla}$ denotes the spatial gradient.
Equation (\ref{Seebeck}) may be viewed as a generalization of the Gibbs-Duhem
relation to the dynamical response of the coupled intensive
variables $(\mu,T)$ by defining the thermodynamic forces acting on the system
as the gradients of the thermodynamic potentials. Each force is conjugated to
a flux which is proportional to the time derivative of the corresponding
extensive parameter, and it follows that forces and fluxes are linearly
combined as proposed by Onsager \cite{Onsager1}. Note that the coefficient $\alpha$, 
which in Eq. (\ref{Seebeck}) is defined locally, is also often expressed as
$\alpha\approx-e\Delta V/\Delta T$, where $\Delta$ denotes the global
difference between two distant points of the system.

These general considerations apply for any system containing free charge carriers. 
As a consequence, for a given temperature difference, the measurement of a thermovoltage 
directly gives an estimation of the average ratio of charge concentrations in separate 
parts of the system. Therefore, any system containing free mobile
carriers may be exhibit a thermoelectric signature. One should note that 
depending on the specific properties of the considered system, this signature may 
intertwine or not with those of other temperature-dependent processes; hence the 
need for specific system-dependent approaches to extract the thermoelectric signal.
In the case of liquids and gels, the studies usually focus on the
thermoelectrophoresis parameters \cite{LiquidThermoelec,GelThermoelec,Majee}. 
For solid state-systems, the conduction is 
ensured by mobile electrons and holes, leading to a possible steady-state
electrical current. For ionic carriers, in liquids and gels, such a
steady-state current may only take place if a redox process occurs at the ends
of the system.

\section{Experiment}
The experiment aims to identify and characterize the thermoelectric response
of a living plant (a bean sprout in the present work). As the response of the plant necessarily 
entails physiologic effects, the measured data is compared to the purely thermoelectric 
response of a control wire. As depicted on Fig.~\ref{fig:figure1}, the temperature of the plant's 
roots is imposed by a thermostatic bath; two thermocouples are used to measure the temperatures of 
the roots and the leaves, and two electrodes are placed respectively in contact with the roots 
and the leaves of the living plant. The roots are bathed in an aqueous dilute KCl solution. 
The electrodes connected to the leaves are made of Ag/AgCl wires. All the temperatures and
voltages are recorded using a Keithley K2700 scanning multimeter and a K182
nanovoltmeter. The thermoelectric control wire (CW) is made of two chromium-nickel-steel 
junctions, with a Seebeck coefficient $\alpha_{\rm junction}=4.6~\mu$V/K; it is placed in the 
same configuration as that of the plant. The recorded voltage at the ends of the control wire describes 
a \emph{pure} thermoelectric response, exempt from any influence of the physiologic process. 

\begin{figure}[th]
\includegraphics*[width=0.5\textwidth]{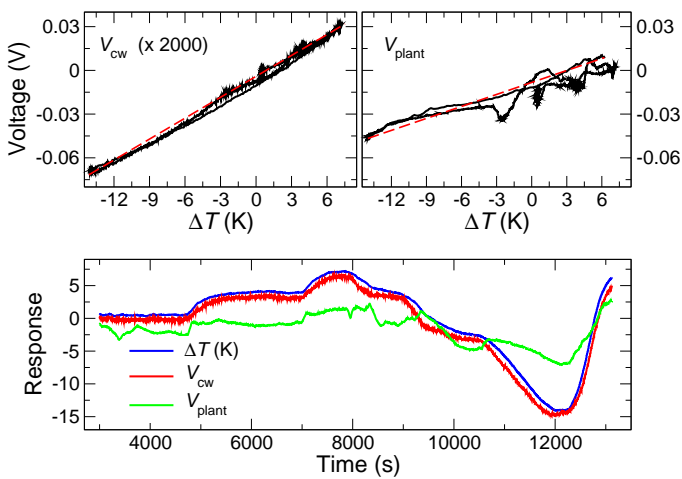} \caption{Experimental data including the 
voltage response of the control wire $V_{\rm cw}$ (upper left panel) and of the plant $V_{\rm plant}$ 
(upper right panel), both submitted to a time-varying temperature difference $\Delta T$, reported on 
the lower panel. The magnitude of measured voltage $V_{\rm cw}$ has been multiplied by a factor 2000 in 
order to report it on the same scale as that of $V_{\rm plant}$. The red-dotted lines indicate the average 
slope of the collected data. In the lower panel, the scale is that of the temperature difference; the 
voltages are also depicted in this lower panel to see how close their time-dependent behaviour is to 
that of the measured temperature difference.}%
\label{fig:Fig2}
\end{figure}

The voltage response of the living plant and that of the control wire to the temperature difference 
$\Delta T$ are both shown on Fig.~\ref{fig:Fig2}, where the time evolution of $\Delta T$ is also displayed, 
illustrating the dynamics of the plant and that of the control wire. These data were recorded over a period  
of four hours for various temperature differences and variation rates. The two upper panels show separately the 
parametric plots characterising the thermoelectric response of the control wire and that of the plant. Notice that 
the response of the control wire is strictly proportional to the temperature difference, as expected for a thermoelectric 
material, with a \emph{passive} Seebeck coefficient. Conversely, the plant exhibits a complex response, which 
includes the effects of its physiological processes thus giving rise to an \emph{active}
Seebeck response. The curves represented on the lower panel are displayed together on the same scale to 
qualitatively show how the responses (voltages $V_{\rm cw}$ and $V_{\rm plant}$) follow the applied 
constraint (temperature difference $\Delta T$). The figures on the $y$-axis correspond to $\Delta T$ in $^{\circ}$C.
The values taken by the voltages $V_{\rm cw}$ and $V_{\rm plant}$ may be determined from the curves using a scaling 
factor that corresponds to the relevant Seebeck coefficient. The control wire shows a faithful homothetic response 
to the temperature difference, with a linear factor of scaling of $4.6~\mu$V/K as
expected for the Seebeck coefficient of the corresponding chromium-nickel-steel thermocouple.
The response of the plant gives an average Seebeck coefficient of 2.5 mV/K. Such a large value  
in a solid-state material would correspond to a system with low electrical conductivity. Here, this
results from ion movements through xylem and phloem saps circulating in opposite directions.

\begin{figure}[th]
\includegraphics*[width=0.5\textwidth]{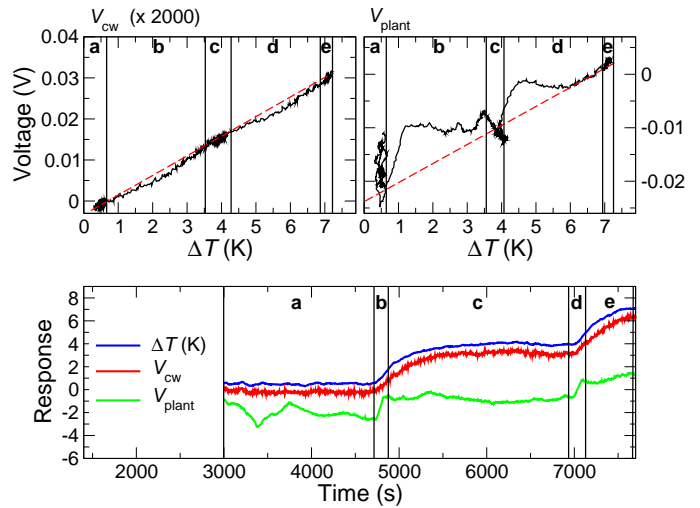}
\caption{First heating sequence. 
The plant's response is strongly influenced by the temperature variation rate ${\rm d}T/{\rm d}t$.}%
\label{fig:Fig3}%
\end{figure}

\begin{figure}[th]
\includegraphics*[width=0.5\textwidth]{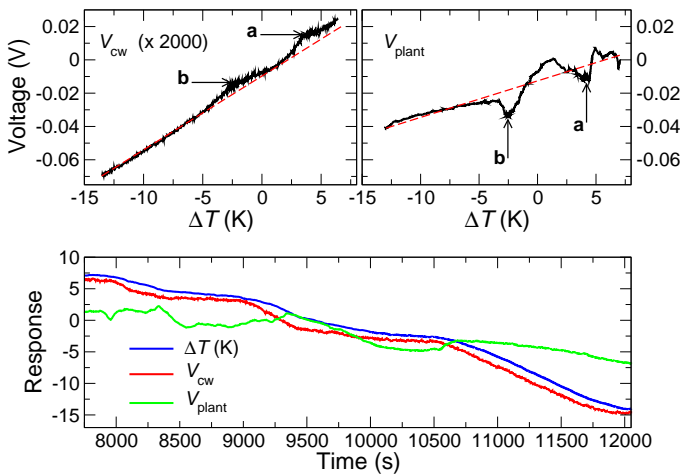}
\caption{Cooling sequence. The response is roughly 
linear but visible discrepancies are due to the plant's physiological processes.}%
\label{fig:Fig4} 
\end{figure}

\begin{figure}[th]
\includegraphics*[width=0.5\textwidth]{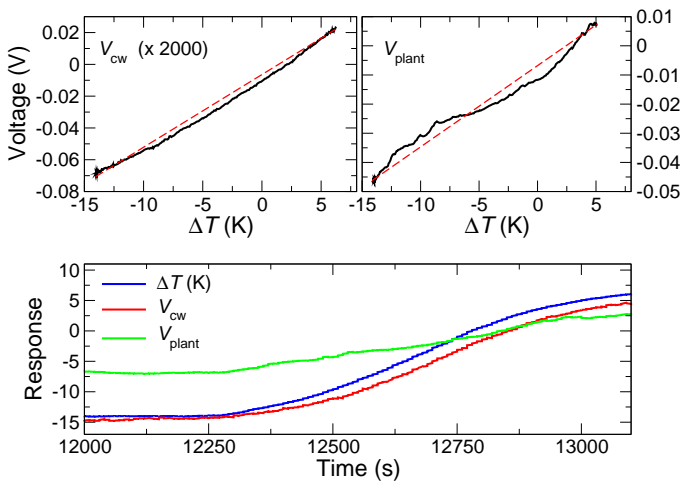}
\caption{Second heating sequence.}%
\label{fig:Fig5}
\end{figure}

To analyze precisely these data, it is convenient to split each panel of Fig.~\ref{fig:Fig2} 
into three parts, which correspond to three distinct time domains. The first and third ones, 
Figs.\ref{fig:Fig3} and \ref{fig:Fig5}, respectively, describes the plant's response when its 
roots are heated; while the second, Fig.\ref{fig:Fig4}, reports the response over the cooling 
sequence.

Depending on the precise thermal condition supported by the plant,
we can identify different thermoelectric responses. Consider the first warming sequence 
of Fig.~\ref{fig:Fig3}. One may identify five events (labelled from a to e).
During the first period (Fig.~\ref{fig:Fig3}a), the temperature difference is
roughly constant. The plant presents a large fluctuating thermoelectric response while
the control wire is characterized by a constant Seebeck coefficient. During the second period, 
shown on Fig.~\ref{fig:Fig3}b, the plant's reaction is essentially sensitive to the
temperature variation rate ${\rm d}T/{\rm d}t$. This sensitivity has already been observed 
in studies devoted to plant sensing \cite{Plieth1,Plieth2}. 
The processes taking place within the zones a and b can be also observed in the c and d zones; 
and when the rate ${\rm d}T/{\rm d}t$ is reduced, we observe a linear response for the plant, 
as shown in the zone e. It thus appears clearly that the passive response, probably mostly due 
to the xylem properties, can be strongly modified by active physiological processes. 
This latter appears to be particularly sensitive to the time derivative of the stress, but not 
so much on the intensity of the stress itself; this observation is perfectly consistent with 
previous findings \cite{Plieth1,Plieth2}. 

We now turn to the cooling sequence, Fig.~\ref{fig:Fig4}. The cooling rate is approximately constant, 
except for the two events when it is clearly larger than the average as indicated with the labels a and b, 
on the upper right panel. These events relate to time delays, also apparent in the measured 
data for the control wire on the upper left panel: these are signatures of the change of 
rate ${\rm d}T/{\rm d}t$. Note that although these may serve as a probe, time delays must be considered with care, 
since in the case of very large time response of a system, they may lead to erroneous interpretations \cite{GelThermoelec}.
The shape of the plant response for events taking place in the zones a and b is very similar to that of Figs.~\ref{fig:Fig3}b 
and \ref{fig:Fig3}d, which confirms the acute sensitivity of the plant to the rate ${\rm d}T/{\rm d}t$. 
This physiological response is the fastest response to a stress that a plant can use, as it has no means to avoid it.

The last sequence depicted on Fig.~\ref{fig:Fig5} concerns the plant's response to warming at a reduced rate. 
It is essentially linear, thus confirming the presence of a threshold rate process for the plant's response to a given stress. 
It also shows unambiguously that thermoelectricity naturally partakes in the complex processes that contribute to a plant's life.

\section{Discussion and concluding remarks}

All the effects discussed in the present work are related to the internal modifications of the ionic
concentrations inside the plant, including the regulation of cellular ion transporters 
allowing ion release into the xylem sap toward the shoots. It is clear that while the response is mainly 
driven by a classical thermoelectric process with a giant Seebeck coefficient, the physiology of the plant 
interferes in a complex way with it. A complete numerical model, including series and parallel assemblies of
living (phloem sieve tube elements, root and endodermis cells) and dead cells (xylem tracheids and vessels), will 
be considered for further investigation of the interplay between physiologic and thermoelectric response to a thermal stress. 
Further, since there is no limitation in the number of thermodynamic potentials which may be coupled \cite{Onsager1}, 
we may account for pressure and derive a complete response of a system to the direct and the cross- effects of
the three thermodynamic potentials, $(P,\mu,T)$. Then the lowering of a local pressure gradient, by the 
coupled effect of the temperature and the electrochemical potential becomes possible. This latter process may 
contribute to the motion of saps in plant, and especially in trees.

The reported measurements are fully scalable from the unique cell size to a complete plant or tree. 
In addition to the stationary response observed here, the voltage fluctuations are also of great interest 
since they permit the study of the threshold of the physiological response through its noisy signals. 
As observed in other systems, the noise response is a very sensitive probe of the emergence of a macroscopic 
response \cite{noise1,noise2}. Taking each cell as a fluctuator, the convolution of the individual responses may lead 
to different signatures depending on the considered scale of the sample. 

\appendix

\section{Plant material} 
Seeds of bush bean (Phaseolus vulgaris L. cv. contender) were sown into pots containing vermiculite as soil. 
The bean seedlings were grown in a growth chamber at $25 \pm 2^{\circ}$C with a cycle of 12 hours of light 
(40 $\mu$mol photon m$^{-2}\cdot$ s$^{-1}$), 12 hours of darkness. The plants were watered at the bottom of 
the pots every three days. The 21 days old plants were removed from the vermiculite. Roots were carefully 
rinsed with water to remove vermiculite particles and then transferred in dilute KCl solution (120 mM) 
for experiments.

\section{Measurement electrodes}
Two electrodes were placed respectively at the surface of the root and the stem of the living plant. The electrodes 
connected to the stems are made of Ag/AgCl wires moistened with 200 mM KCl and wrapped in cellophane to provide 
appropriate contact with the plant surface.

\end{document}